# Dynamic control of polarization conversion based on borophene nanostructures in optical communication bands

Xinyang Wang,[1, 2] Qi Lin,[1, 3] Lingling Wang[3] and Guidong Liu[1, 3, *]

[1]*School of Physics and Optoelectronics, Xiangtan University, Xiangtan 411105, China*

[2]*Hunan Engineering Laboratory for Microelectronics, Optoelectronics and System on a Chip, Xiangtan University, Xiangtan 411105, China*

[3]*School of Physics and Electronics, Hunan University, Changsha 410082, China*

*gdliu@xtu.edu.cn

**Abstract:** Polarized light has various potential applications in the communication bands, including optical communication, polarization imaging, quantum emission, and quantum communication. However, optimizing polarization control requires continuous improvements in areas such as dynamic tunability, materials, and efficiency. In this work, we propose a borophene-based structure capable of converting linearly polarized light into arbitrarily polarized light through the coherent excitation of localized surface plasmons (LSPs) in optical communication band. Furthermore, a double-layer borophene structure can be achieved by placing a second borophene array at the top of the first one with a 90 ° relative rotation of their crystalline plane. The rotation direction of the polarization state of the reflected light can be switched by independently controlling the carrier concentration of the two-layer borophene. Finally, a dipole source is used to realize the emission of polarized light, which is two orders of magnitude higher than the emission rate in free space, and the polarization state can be dynamically controlled by manipulating the carrier concentration. Our study is simple and compact, with potential applications in the fields of polarizers, polarization detectors, and quantum emitters.

1. Introduction

Polarization is one of the intrinsic properties of electromagnetic waves, which represents the property of the change in direction of an electromagnetic vector in space [1], including three polarization states: linear polarization light (LPL), elliptical polarization light (EPL) and circular polarization light (CPL). In the field of communication and sensing, compared with LPL, CPL makes the light resistant to environmental changes and ignores the effects of scattering and diffraction [2-4]. It is difficult to generate CPL directly, but the LPL can be converted into CPL by adjusting the electromagnetic amplitude and phase between the two orthogonal electric field components [5]. Metamaterials can flexibly manipulate the scattering amplitude, phase and polarization of light, and can theoretically shape the wavefront of light into any desired shape. Early studies of polarization conversion have shown that metamaterials composed of noble metals [6, 7] or all-dielectric materials [7, 8] can control the polarization state of light. However, in order to dynamically control the polarization state or operating wavelength, the structural parameters must be redesigned.

Two-dimensional (2D) materials such as graphene generally support plasmonic modes, and their optical response can be controlled by adjusting their carrier concentration through electrostatic gating or chemical doping. For instance, graphene [9-13] and black phosphorus (BP) [14] can manipulate the amplitude and phase of light through electrical tunability, which facilitates rapid and flexible dynamic adjustment of polarization conversion. In fact, the study of dynamic tunable polarization conversion is of great commercial value in the communication bands, which can be widely used in polarization imaging [15], astronomical observation [16], optical communication [17] and other fields. For 2D materials such as graphene and BP, their research on polarization covers terahertz [18-20] and gigahertz [21] bands. However, graphene and BP do not support plasmon modes in the optical communication band. Recently, phase change materials such as GST have been introduced into nanostructures [22, 23] for achieving tunable polarization conversion devices in 1550 nm. However, it is achieved by heat treatment of GST hybrid nanostructures, which inevitably leads to a relatively long tuning time. As an emerging 2D material, borophene exhibits anisotropic plasmonic response in the near-infrared [24], and its carrier concentration can be regulated by electrostatic gating or chemical doping [25], which makes borophene an ideal material for dynamically tunable polarization conversion in the communication band.

In this work, we propose a borophene based nanostructure to achieve the desired polarization control. The numerical results show that the adjustment of the carrier concentration of borophene can manipulate the spectral shift of coherently excited LSPs in the communication band, so as to realize the conversion from LPL to arbitrary polarized light, and CPL can also be converted to LPL. A double-layer borophene structure can be achieved by placing a second borophene array at the bottom of the first one with a 90° relative rotation of their crystalline plane. The rotation direction of the polarization state of the reflected light can be switched by independently controlling the carrier concentration of the two-layer borophene. Finally, a dipole source is used to realize the emission of polarized light, which is two orders of magnitude higher than the emission rate in free space, and the polarization state can be dynamically controlled by controlling the carrier concentration. Our research provides a good opportunity for the development of tunable polarization conversion in the communication bands.

## 2. Proposed design and principle of operation

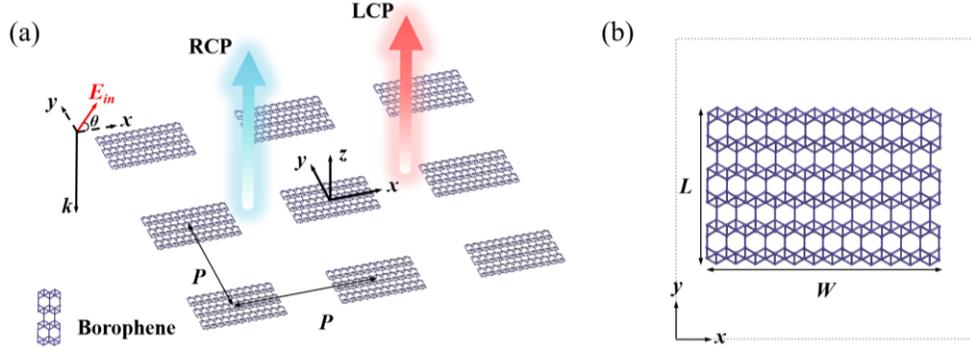

Fig. 1. (a) Structure diagram of borophene array. (b) Top view of the structure in a unit cell.

The proposed structure is shown as Fig. 1(a), where the LPL normal incident in the opposite direction along the z-axes. The detailed geometry parameters: $P = 60$ nm, $W = 50$ nm, $L = 35.5$ nm, as shown in Fig. 1(b). The finite-difference time-domain (FDTD) method is used to investigate the electromagnetic properties of the structure. The grids in both the x- and y-directions are set to 0.5 nm to ensure accuracy. The x- and y-directions use the periodic boundary condition, while the z-direction uses the perfectly matched layers (PMLs) boundary. The refractive index of the background is set to 1 for the sake of simplicity. The Drude model can be used to describe the conductivity of borophene [25, 26]:

$$\sigma_{ij} = \frac{iD_j}{\pi\left(\omega + \frac{i}{\tau}\right)}, \quad D_j = \frac{\pi e^2 n}{m_j}. \tag{1}$$

Where $e$, $\omega$, $n$ and $\tau$ represent the electron charge, frequency, carrier concentration and average free time of electrons, respectively. The Drude weight and effective electronic mass is given by $D_j$ and $m_j$, respectively; the subscripts $(i, j)$ represent the optical axes of borophene in the x- and y-directions; the electronic mass is marked as $m_0$, the effective electronic masses $m_x$ and $m_y$ are $1.4m_0$ and $3.4m_0$, respectively.

$$\varepsilon_{jj} = \varepsilon_r + \frac{i\sigma_{jj}}{\varepsilon_0 \omega d} \tag{2}$$

Where $\varepsilon_0$ and $\varepsilon_r$ are the vacuum permittivity and the relative permittivity, respectively. $d = 0.3$ nm is the thickness of borophene.

First of all, the LPL is used as excitation source. As shown in Fig. 2(a), the $\theta$ represents the angle between the direction of incident polarization and the x-axes, and the electric field vector $E_{inc}$ of the LPL source is decomposed into $E_x$ and $E_y$ along the coordinate axes, where $E_x = E_{inc}\cos\theta$ and $E_y = E_{inc}\sin\theta$. For a planar system, the polarization of light can be described by a 2 × 2 Jones matrix with four complex elements [27, 28]. Reflective fields $E_{\text{ref}}$ can be described as:

$$\begin{pmatrix} E_{x,\text{ref}} \\ E_{y,\text{ref}} \end{pmatrix} = \begin{pmatrix} r_{xx} & r_{xy} \\ r_{yx} & r_{yy} \end{pmatrix} \begin{pmatrix} E_{x,\text{inc}} \\ E_{y,\text{inc}} \end{pmatrix} \qquad (3)$$

$r_{xx}$, $r_{yy}$, $r_{xy}$, and $r_{yx}$ are the reflected amplitudes, and the $x$ and $y$ subscripts represent different polarization directions. Here $r_{xy} = r_{yx} = 0$, so we can simplify Eq. (3) to:

$$\boldsymbol{E}_{\text{ref}} = r_{xx} E_{x,\text{inc}} \boldsymbol{x} + r_{yy} E_{y,\text{inc}} \boldsymbol{y} \qquad (4)$$

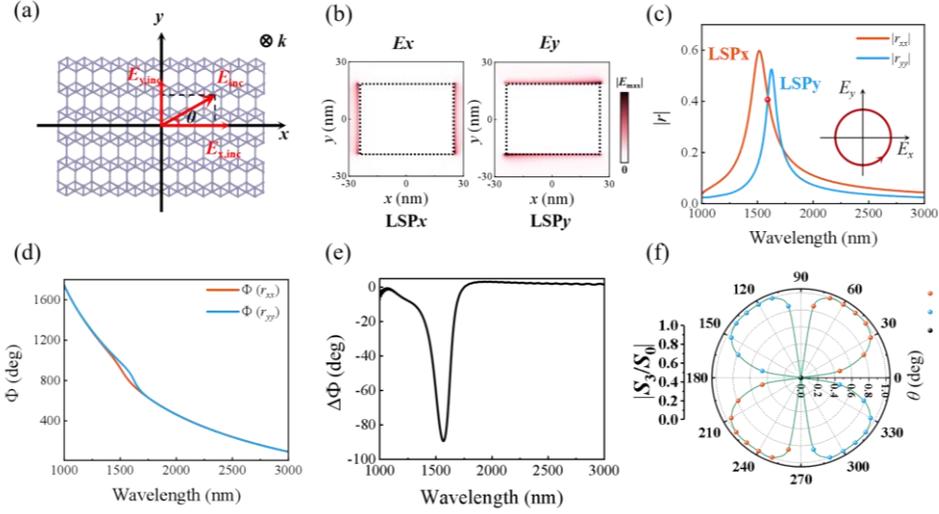

Fig. 2. (a) Schematic diagram of the LP incident light polarization plane. (b) The electric field distribution $E_x$ and $E_y$ on the $x$-$y$ plane at 1550 nm. (c) The amplitudes of reflection coefficients ($r_{xx}$, $r_{yy}$). The inset shows the far-field polarization ellipse at the red dot marker. (d) Phase spectra. (e) Phase difference as a function of wavelength. (f) The Stokes parameter map with different polarization angles.

Unless otherwise stated, the carrier concentration of the selected borophene is $6.3 \times 10^{19}$ m$^{-2}$, and $\theta$ is 45°. Figure 2(b) displays the electric field distribution of $E_x$ and $E_y$, indicating the excitation of two mutually orthogonal plasmon modes. The degeneracy of the two orthogonal dipole modes can be attributed to the anisotropic conductivity of borophene and the relative size of its length and width. The optical response of the two modes is characterized by different resonance wavelengths and intersecting reflection spectra, as shown in the red dot in Fig. 2(c). As shown in Fig. 2(d), the phase parameters $\Phi(r_{xx})$ and $\Phi(r_{yy})$ of the reflected amplitude at the corresponding wavelength obtain a large phase difference, and the phase difference is defined as $\Delta\Phi = \Phi(r_{xx}) - \Phi(r_{yy})$. To more clearly represent the optical response of the proposed structure, we introduce

four Stokes parameters to describe the polarization state of the reflected light:

$$S_0 = |r_{xx}|^2 + |r_{yy}|^2$$
$$S_1 = |r_{xx}|^2 - |r_{yy}|^2$$
$$S_2 = 2|r_{xx}||r_{yy}|\cos\Delta\Phi$$
$$S_3 = -2|r_{xx}||r_{yy}|\sin\Delta\Phi$$

(5)

Here, parameter $S_1 = S_0$ (-$S_0$) corresponds a perfect linear polarized light with the electric field vector along the $x$-axes ($y$-axes); parameter $S_2 = S_0$ (-$S_0$) corresponds a perfect linear polarized light with the angle between the electric field vector and the $x$-axes is 45° (-45°); parameter $S_3 = S_0$ (-$S_0$) corresponds a perfect circularly right handed (left handed) polarized light.

As shown in Fig. 2(e), the phase difference reaches 90° near 1550 nm, which satisfies the necessary condition for forming perfect CPL polarization according to Eq. (5). To investigate how the polarization angle $\theta$ of the incident light affects the polarization state of the reflected light, the $\theta$ is varied and the ellipticity parameters of the reflected light is plotted in Fig. 2(f). The polar diameter and polar angle of the polar coordinates represent the ellipticity parameters $S_3/S_0$ of the reflected light and the angle $\theta$, respectively. The reflected polarization states left-handed (LH), right-handed (RH) and LP are represented by orange, blue and black dots, respectively. When $\theta$ is in the first or third (second or fourth) quadrants, LSP$x$ will be ahead of LSP$y$ $\pi$ (-$\pi$) phase difference, resulting in the conversion of LP to LH (RH) polarized light. We note that when looking against the direction of light propagation, the composite vector rotates clockwise, the polarized light is RH and vice versa. When $\theta$ coincides with the coordinate axes, there is no coherence between the polarized dipole modes in the $x$- and $y$-directions. Hence, no polarization conversion occurs.

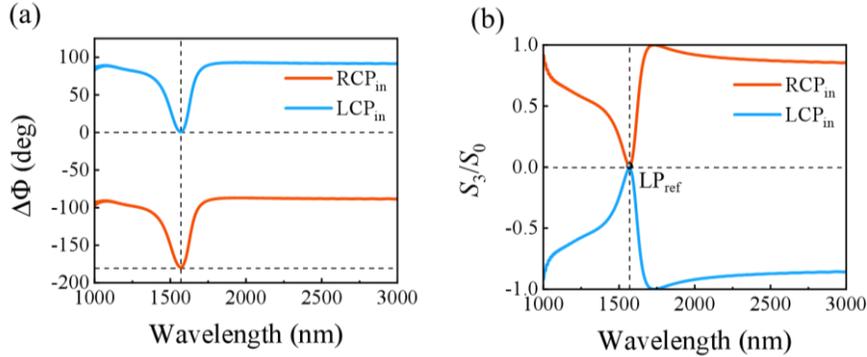

Fig. 3. (a) The phase difference of the reflection amplitude $r_{xx}$, $r_{yy}$ as a function of wavelength. (b) Ellipticity parameters $S_3/S_0$ curve with LCP and RCP incident light.

In the same way, it is also possible to reverse the optical conversion from CPL to LPL, as shown in Fig. 3. In this case, the geometric parameters are the same as those in Fig. 2. The LCP or RCP light normal incidence along the negative $z$-axes direction, and the phase difference $\Delta\Phi$ is shown in Fig. 3(a). When the resonance position is near 1550 nm, the phase difference reaches 0 °(-180 °) under the LCP (RCP) light, and the reflected light is LP. Figure 3(b) shows the ellipticity parameter ($S_3/S_0$) of the reflected light as a function of wavelength under the incident light of LCP (RCP), where the orange and blue lines represent the incident light as RCP and LCP, respectively. In the vicinity of 1550 nm (vertical line), the $S_3/S_0$ value of the reflected light reaches 0, that is, linear polarization. It can be seen that the reverse operation from circular polarization to linear polarization is also feasible.

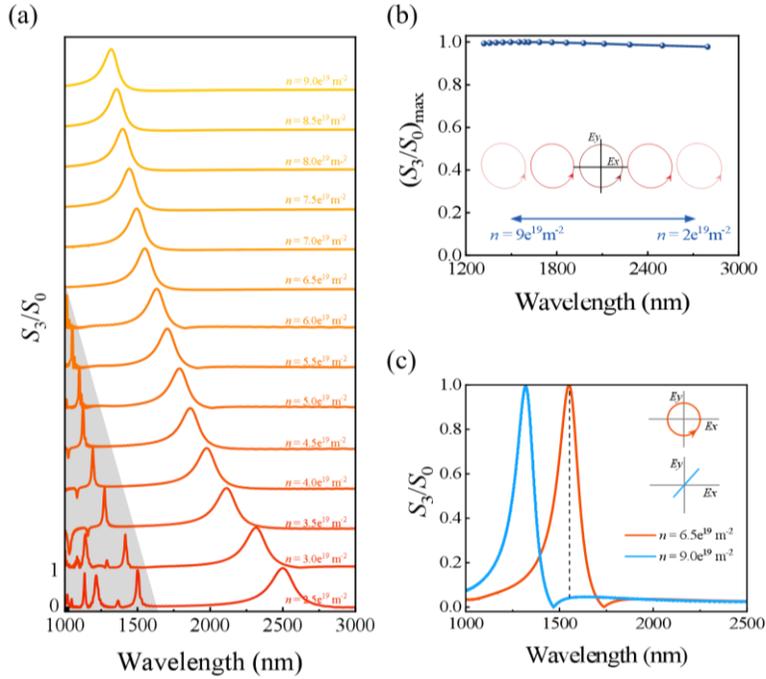

Fig. 4. (a) $S_3/S_0$ spectra with varied $n$. (b) Extract the maximum $S_3/S_0$ for different $n$ in (a). The inset shows the variation trend of the maximum polarization ellipse. (c) Ellipticity parameters $S_3/S_0$ curve with different carrier concentrations. The inset shows the polarized ellipse at the location of the dashed line.

Next, the control of the operating wavelength by dynamically modulating the carrier concentration of the borophene is investigated. In the following example, in general case that the polarization direction of the incident light is $\theta = 45°$, i.e., the amplitude weights of $r_{xx}$ and $r_{yy}$ are equal. Figure 4(a) the dependence of ellipticity on the carrier concentration of borophene. As the carrier concentration increases from $2.5 \times 10^{19}$ m$^{-2}$ to

$9.0 \times 10^{19}$ m$^{-2}$, the resonance gradually blueshifts from 2501 nm to 1320 nm, the modulation range is close to 1200 nm. The sharp spectral line in the shadow is high-order plasmon resonance, which is not concerned in this study. With the increase of carrier concentration, the resonance wavelength gradually blue-shifted, and the maximum polarization ellipticity of the reflected light remained almost unchanged at 1, as shown in Fig. 4(b). On the center wavelength of the third communication window, 1550 nm, when the carrier concentration is $6.5 \times 10^{19}$ m$^{-2}$, the value of $S_3/S_0$ is 1, indicating that the polarization state reflected light is circularly polarized. As the carrier concentration increases to $9.0 \times 10^{19}$ m$^{-2}$, it becomes linear, as indicated by a value close to 0 for $S_3/S_0$. It should be noted that the value of $S_3/S_0$ can be continuously tuned between 0 and 1 not only at the wavelength of 1550 nm but also at other resonance wavelengths by modulating the carrier concentration of borophene.

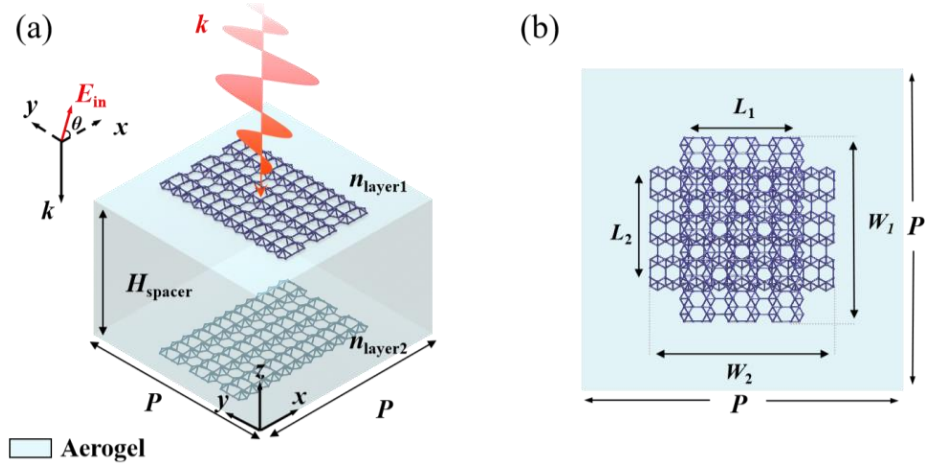

Fig. 5. (a) The structure diagram of the double-layer borophene. (b) Top view of the structure.

In addition to controlling polarization conversion and ellipticity, another desirable feature is the ability to switch handedness between RCP and LCP. In most cases, these exhibits diametrically opposed optical responses to the same structure. Dynamic conversion between LCP and RCP is necessary. As shown in Fig. 5(a), a double-layer borophene structure can be achieved by placing a second borophene array at the top of the first one with a 90° relative rotation of their crystalline plane. The carrier concentrations of the upper and lower layers of borophene are represented by $n_{layer1}$ and $n_{layer2}$, respectively. The distance between them is given by $H_{spacer}$. The LPL source is incident in the -z direction. For simplicity, we set the intermediate dielectric layer as an aerogel with a refractive index ($n_{aerogel} = 1.08$) close to air as mechanical support, and the thickness of the aerogel is $H_{spacer} = 120$. As shown in Fig. 5(b), $L_1$, $L_2$ and $W_1$, $W_2$ represent the length and width of the upper and lower layers of borophene, respectively. Structural parameters is $W_1 = W_2 = W = 50$ nm, $L_1 = L_2 = L = 35.5$ nm and $P = 60$ nm. The electron effective mass of the top layer in the $x$- ($y$-) direction corresponds to that of the lower layer in the $y$- ($x$-) direction. Therefore, compared with the aforementioned single-layer structure, the LSP$x$ (LSP$y$) of the lower layer of borophene is the original LSP$y$ (LSP$x$), and the phase difference sign of the reflected light in the $x$- and $y$-directions is opposite to the upper layer of borophene.

By adjusting the carrier concentration independently, the upper layer of borophene is in a resonant state, while the lower layer of is in a nearly non-resonant state. The overall optical response is similar to that shown in Fig. 2, that is, the LSP at a shorter wavelength provides the dipole moment in the $x$ direction, while the LSP at a longer wavelength provides the dipole moment in the $y$ direction. The opposite results occur when the two layers are rotated 90° against each other. The same result can also be achieved by exchanging the carrier concentration of two layers of borophene.

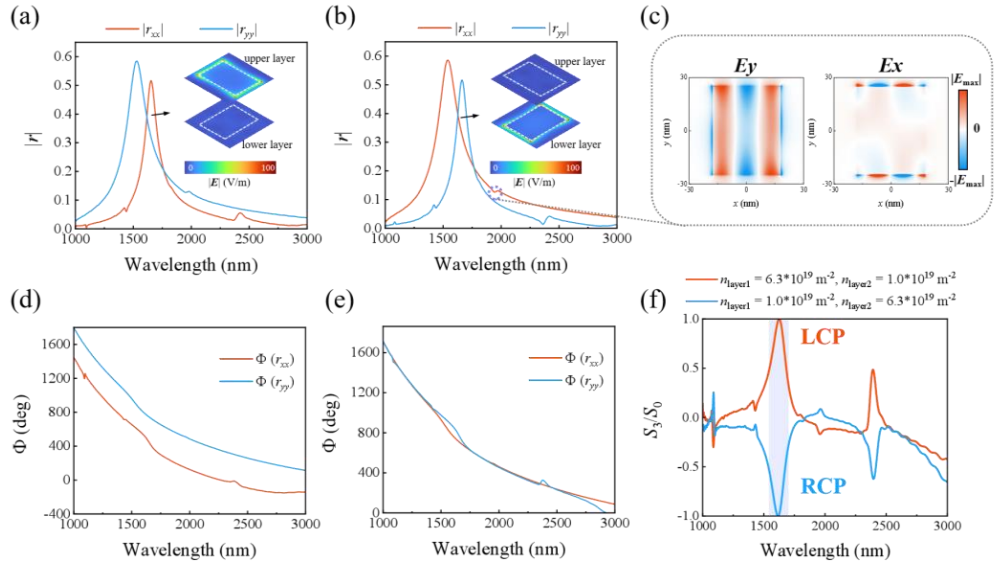

Fig. 6. The reflected amplitudes at (a) $n_{layer1} = 6.3 \times 10^{19}$ m$^{-2}$, $n_{layer2} = 1.0 \times 10^{19}$ m$^{-2}$ and (b) $n_{layer1} = 1.0 \times 10^{19}$ m$^{-2}$, $n_{layer2} = 6.3 \times 10^{19}$ m$^{-2}$, respectively. The insets show the electric field distribution of the upper and lower layers of borophene. (c) The electric field distribution $E_x$ and $E_y$ of the resonance at the circle marker in (b). (d) and (e) correspond to the phase spectra of (a) and (b), respectively. (f) Ellipticity parameters $S_3/S_0$ curve in both cases.

In Fig. 6(a), by selecting $n_{layer1} = 6.3 \times 10^{19}$ m$^{-2}$, and at 1586 nm, the LSPs mode of borophene is excited. By selecting $n_{layer2} = 1.0 \times 10^{19}$ m$^{-2}$, there is no mode excitation of the lower borophene at 1586 nm, which can be observed from the inset. The corresponding phase spectrum is shown in Fig. 6(d). The phase difference $\Delta\Phi$ is -270° at the resonance position with equal reflection amplitude, and the reflected light is LCP. When the carrier concentration of the upper and lower layers of borophene is exchanged, the phase difference $\Delta\Phi$ becomes 90°, and the reflected light is RCP, as shown in Figs. 6(b) and (e). The tiny resonance near 1963 nm in Fig. 6(b) is the higher-order mode of the plasmons, which can be identified from the electric field distribution in Fig. 6(c). Figure 6(f) plots the ellipticity parameters in two cases, where the shadow part shows the inversion of the $S_3/S_0$ symbol near 1586 nm, proving the transformation of the rotation of the reflected light. With further adjustment of the carrier concentrations $n_{layer1}$ and $n_{layer2}$, all possible polarization conversion can be achieved, similar to the case of the single-layer borophene shown in Fig. 2.

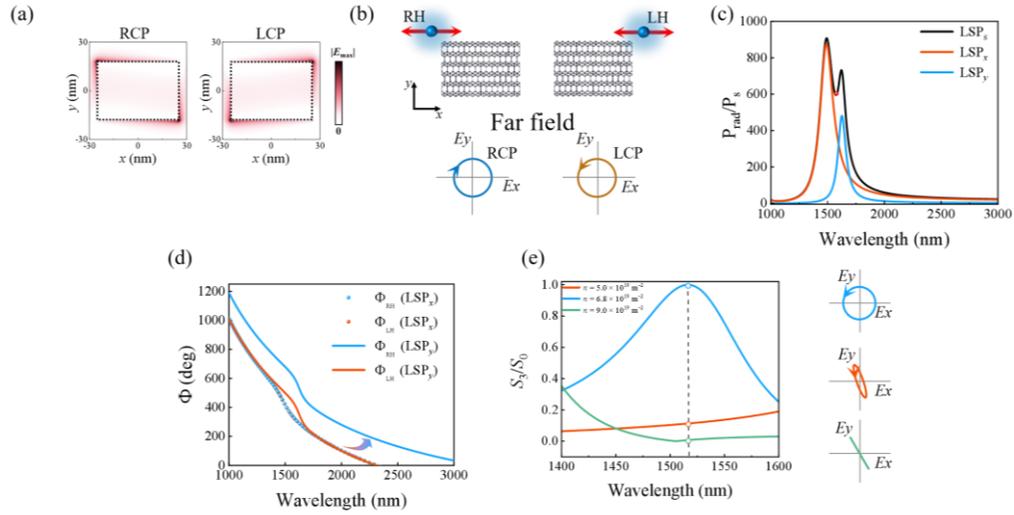

Fig. 7. (a) Near-field distribution plotted with far-field excitation of LCP and RCP plane light, respectively. (b) The schematics show the dipole positions for exciting LCP (yellow) and RCP (blue) emission light, respectively. (c) Emission enhanced spectrum. (d) The phase spectra of LCP and RCP emitted light. (e) Ellipticity parameters $S_3/S_0$ curve with different carrier concentrations. The inset shows the far-field polarized ellipse.

Recent studies of achiral nanostructures that support external chirality have shown that CPL emission can be achieved when the dipole source is located at a specified hotspot [29, 30]. Combined with this principle, it is also possible to actively control the polarization state of the emitted light. In this case, the geometric parameters are the same as those in Fig. 2. The near-field distribution under RCP and LCP illumination is shown in Fig. 7(a), where the combined action of LSP$x$ and LSP$y$ under CPL excitation leads to the formation of a diagonally distributed plasmon mode. According to the reciprocity, the dipole source will be placed near the near-field hotspot around the borophene nanoantenna. It will reproduce the observed spatial characteristics of the CPL far-field. As shown in Fig. 7(b), the dipole source is all $x$-polarized, which are set at 1.5 nm in the $x$-direction and 0.5 nm in the $y$-direction from the left and right vertices of the nanoantenna, and the LSP$x$ and LSP$y$ are excited by the local $x$-polarization components of the two LSPs near the dipole source. The following illustration shows the polarization ellipse of the transmitted light when the dipole source is set at different positions, where the arrow indicates the rotation direction of the far-field polarization ellipse. The reason for this emission of an opposite-rotating light is that the phase difference between LSP$x$ and LSP$y$ increases by $\pi$ when the dipole source is moved from the top right corner to its top left corner of the nanoantenna, as shown in Fig. 7(d). Figure. 7(c) shows the emission enhancement spectrum, where $P_{rad}$ is the emission power and $P_s$ is the emission power of the same dipole source in free space. Borophene nanostructures can provide additional radiation pathways to enhance emission through the Purcell effect, and the $P_{rad}$ is two orders of magnitude higher than the $P_s$. As shown in Fig. 7(e), the dipole source is placed in the upper right corner of the structure. The carrier concentration of borophene is varied to $n = 5.0 \times 10^{19}$ m$^{-2}$, $n = 6.5 \times 10^{19}$ m$^{-2}$ and $n = 9.0 \times 10^{19}$ m$^{-2}$. The carrier concentration increases with the blue shift of the LSPs mode. Therefore, at 1517 nm, elliptically polarized light, circularly polarized light and linearly polarized light are emitted respectively. It can be seen that the ellipticity of the emitted light can be actively controlled by manipulating the carrier concentration. We demonstrate this tunable polarization state of far-field emission, which can provide flexible applications in the communication band, such as quantum emitters with tunable polarization states.

**3.** Conclusion

In summary, we propose a borophene-based structure capable of converting linearly polarized light into arbitrarily polarized light through the coherent excitation of LSPs in optical communication band. In addition, a double-layer borophene structure can be achieved by placing a second borophene array at the top of the first one with a 90 ° relative rotation of their crystalline plane. The direction of rotation of the reflected polarization state can be switched by independently controlling the carrier concentrations in the upper and lower layers. Finally, we use a dipole source to excite LSPs at selected hotspot to induce the emission of CPL, and increase the emission rate by two orders of magnitude. The polarization state of the emitted light can be dynamically tuned by manipulating the carrier concentration of borophene. Our research provides a flexible solution in the near-infrared spectral range to design applications with highly controllable polarization states, such as dynamically tunable nanoscale polarizers.

Conflict of interest

There are no conflicts to declare.

Acknowledgements.

This work is supported by the Scientific Research Foundation of Hunan Provincial Education Department (22B0105), Hunan Provincial Natural Science Foundation of China (2021JJ40523, 2020JJ5551), and the National Natural Science Foundation of China (62205278, 11947062).